\documentclass[12pt]{article}
\usepackage{epsfig,psfrag,harvard}
\usepackage{amsmath}

\citationmode{abbr}

\citationmode{abbr}

\def\sign{{\rm sign}}

\begin{document}
\title{Sparse inverse covariance estimation with the lasso}
\author{
{\sc Jerome Friedman}
\thanks{Dept. of Statistics, Stanford Univ., CA
94305, jhf@stanford.edu}\\
{\sc Trevor Hastie}
\thanks{Depts. of Statistics, and Health, Research \&
  Policy, Stanford Univ., CA
94305, hastie@stanford.edu}\\
and
{\sc Robert Tibshirani}\thanks{Depts. of Health, Research \&
  Policy, and Statistics,
    Stanford Univ, tibs@stanford.edu}
}

\maketitle
\begin{abstract}
We consider the problem of estimating sparse graphs by  
a lasso penalty applied to the inverse covariance matrix. 
Using a coordinate descent procedure for the lasso,
we develop a simple algorithm that is remarkably fast: in the worst
cases, it solves a 1000 node problem ($\sim 500,000$ parameters) in about
a minute, and is 50 to 2000 times faster than competing methods.
 It also provides a conceptual link between the exact problem
and the approximation suggested by \citeasnoun{MB2006}.
We illustrate the method on some cell-signaling data from proteomics.
\end{abstract}

\section{Introduction}
In recent years a number of authors have proposed the estimation of
sparse undirected graphical models through the use of  $L_1$ 
(lasso) regularization. The basic model  for continuous data assumes
that the observations have a multivariate Gaussian distribution with mean $\mu$ and covariance matrix $\Sigma$.
If the $ij$th  component of $\Sigma^{-1}$ is zero, then variables $i$ and $j$
are conditionally independent, given the other variables.
Thus it makes sense to impose an $L_1$ penalty for the estimation of $\Sigma^{-1}$.

\citeasnoun{MB2006} take a simple  approach to this problem:  they estimate a
sparse graphical model by fitting a lasso model to each variable,
using the others as predictors. The component $\hat\Sigma^{-1}_{ij}$ is then estimated
to be non-zero if  either  the estimated coefficient of variable $i$ on
$j$, or the estimated coefficient of variable $j$ on
$i$, is non-zero (alternatively they use an AND rule).
They show that asymptotically,  this consistently estimates the set
of non-zero elements of $\Sigma^{-1}$.

Other authors have proposed algorithms for the exact
maximization of the $L_1$-penalized  log-likelihood; both
\citeasnoun{YL2007a} and \citeasnoun{BGA2007} adapt interior point
optimization methods for the solution to this problem.
Both papers also establish that the  simpler approach of \citeasnoun{MB2006}
can be viewed as an approximation to the exact problem.

We use the development in \citeasnoun{BGA2007} as a launching point,
and propose a simple, lasso-style algorithm for the exact problem.
This new procedure is  extremely simple,
and is substantially faster than the interior point
approach in our tests.
It also bridges the ``conceptual gap'' between the 
\citeasnoun{MB2006} proposal and the exact problem.
\section{The proposed method}
Suppose we have
  $N$ multivariate normal observations of dimension $p$, with mean $\mu$
and covariance $\Sigma$.
Following \citeasnoun{BGA2007}, let $\Theta=\Sigma^{-1}$, and let
$S$ be the
empirical covariance matrix,
the problem is to maximize  the log-likelihood
\begin{eqnarray}
\log \det \Theta-{\rm tr}(S\Theta)-\rho||\Theta||_1,
\label{one}
\end{eqnarray}
where ${\rm tr}$ denotes the trace and $||\Theta||_1$ is the $L_1$ norm--- the
 sum of the
absolute values of the elements of $\Sigma^{-1}$.
Expression~(\ref{one}) is the Gaussian log-likelihood of the data, partially maximized 
with respect to the mean parameter $\mu$.
\citeasnoun{YL2007a} solve this problem using the interior
point method for  the ``maxdet''  problem, proposed by \citeasnoun{VBW1998}.
\citeasnoun{BGA2007}  develop a different framework for the optimization, which 
was the impetus for our work.

\citeasnoun{BGA2007}  show that the problem (\ref{one}) is convex
and consider estimation of $\Sigma$ (rather than 
$\Sigma^{-1}$), as follows.
Let $W$ be the estimate of $\Sigma$.
They show that one can solve the problem by optimizing over each  row
and corresponding column of $W$ in a block coordinate descent fashion.
 Partitioning $W$ and $S$
\begin{equation}
\label{eq:1}
W=
\begin{pmatrix}
W_{11} & w_{12} \\
w_{12}^T & w_{22}
\end{pmatrix},
\;\;\;
S=\begin{pmatrix}
S_{11} & s_{12} \\
s_{12}^T  &s_{22}
\end{pmatrix},
\end{equation}
they show that the solution for $w_{12}$ satisfies 
\begin{eqnarray}
\hat w_{12}={\rm argmin}_y\{y^TW_{11}^{-1}y: ||y-s_{12}||_\infty \leq \rho \}.
\label{two}
\end{eqnarray}
This is a box-constrained quadratic program which they solve using an interior point procedure.
Permuting the rows and columns so the target column is always the
last, they solve a problem like (\ref{two}) for each column, updating
their estimate of $W$ after each stage. This is repeated until convergence.
Using convex duality,
\citeasnoun{BGA2007} go on to show that
(\ref{two}) is equivalent to the dual problem
\begin{eqnarray}
{\rm min}_\beta ||W_{11}^{1/2}\beta-b||^2 + \rho ||\beta||_1,
\label{three}
\end{eqnarray} 
where $b=W_{11}^{-1/2}s_{12}/2$.  This expression is the basis for our approach.

First we note that it is easy to   verify  the equivalence between the solutions to
(\ref{one}) and (\ref{three}) directly.
The sub-gradient equation for  maximization of the log-likelihood (\ref{one})
is 
\begin{equation}
W-S-\rho\cdot \Gamma = 0,
\label{gradd}
\end{equation}
using the fact that the derivative of  $\log\det \Theta$ equals $\Theta^{-1}=W$,
given in e.g \citeasnoun{BV2004}, page 641. Here $\Gamma_{ij}\in{\rm
  sign}(\Theta_{ij})$; i.e. $\Gamma_{ij}={\rm sign}(\Theta_{ij})$ if
$\Theta_{ij}\neq 0$, else $\Gamma_{ij}\in[-1,1]$ if $\Theta_{ij}=0$.

Now the upper right block  of  equation (\ref{gradd}) is
\begin{eqnarray}
w_{12}-s_{12}-\rho\cdot \gamma_{12} = 0,
\label{grad1}
\end{eqnarray}
using the same sub-matrix notation as in (\ref{eq:1}).

On the other hand, the sub-gradient equation from (\ref{three}) works out to be
\begin{eqnarray}
2W_{11}\beta-s_{12}+\rho\cdot \nu = 0,
\label{grad2}
\end{eqnarray}
where $\nu\in  {\rm sign}(\beta)$ element-wise.

Now suppose $(W,\Gamma)$ solves (\ref{gradd}), and hence
$(w_{12},\gamma_{12})$ solves (\ref{grad1}). Then
$\beta=\frac12W_{11}^{-1}w_{12}$ and $\nu=-\gamma_{12}$ solves
(\ref{grad2}).
The equivalence of the first two terms is obvious. For the sign terms,
since $W_{11}\theta_{12}+w_{12}\theta_{22}=0$, we have that
$\theta_{12}=-\theta_{22}W_{11}^{-1}w_{12}$ (partitioned-inverse formula). Since
$\theta_{22}>0$, then  $\sign(\theta_{12})=-\sign(W_{11}^{-1}w_{12})=-\sign(\beta)$.

Now to the main point of this paper.  Problem~(\ref{three}) looks
like a lasso ($L_1$-regularized) least squares problem.  In fact if
$W_{11}=S_{11}$, then the solutions $\hat\beta$ are easily seen to
equal one-half of the lasso estimates for the $p$th variable on the
others, and hence related to the \citeasnoun{MB2006} proposal.  As
pointed out by \citeasnoun{BGA2007}, $W_{11} \neq S_{11}$ in general
and hence the \citeasnoun{MB2006} approach does not yield the maximum likelihood
estimator.  They point out that their block-wise interior-point
procedure is equivalent to recursively solving and updating the lasso
problem (\ref{three}), but do not pursue this approach.  We do, to
great advantage, because fast coordinate descent algorithms
\cite{FHT2007} make solution of the lasso problem very attractive.

% But we can use a modified version the lasso approach    to obtain the
% actual maximum likelihood estimates of the  problem (\ref{three}).
In terms of inner products, the
usual lasso estimates for the $p$th variable on the others take as input the data
$S_{11}$ and $s_{12}$.
To solve (\ref{three}) we instead use $W_{11}$ and $s_{12}$,
where $W_{11}$ is our current estimate of the upper block of $W$.
We then update $w$ and cycle through all of the variables until convergence.

Note that from (\ref{gradd}), the solution $w_{ii}=s_{ii}+\rho$ for
all $i$, since $\theta_{ii}>0$, and hence $\Gamma_{ii}=1$.
Here is our algorithm in detail:

\bigskip

\centerline{\em Covariance Lasso Algorithm}
\begin{enumerate}
\item Start with $W= S+\rho I$. The diagonal of $W$ remains unchanged in 
what follows.
\item For each $j=1,2,\ldots p, 1,2,\ldots p, \ldots$,  solve the lasso problem
(\ref{three}), which takes as input the inner products $W_{11}$ and $s_{12}$.
This gives a $p-1$ vector solution $\hat\beta$.
Fill in the corresponding row and column of $W$ using 
$w=2W_{11} \hat\beta$.
\item Continue until convergence
\end{enumerate}

\bigskip

Note again that each step in step (2) implies a permutation of the rows and
columns to make the target column the last. 
The lasso problem in step (2) above
can be efficiently solved by coordinate descent (\citeasnoun{FHT2007},\citeasnoun{WL2007a}).
Here are the details.
 Letting $V=W_{11}$, then the
update has the form
\begin{eqnarray}
\hat\beta_j \leftarrow S(s_{12j}-2\sum_{k\neq j} V_{kj}\hat\beta_k, \rho)/(2V_{jj})
\end{eqnarray}
for $j=1,2,\ldots p, j=1,2,\ldots p, \ldots$,
where $S$ is the soft-threshold operator:
\begin{eqnarray}
S(x,t)= {\rm sign}(x)(|x|-t)_+.
\end{eqnarray}
We cycle through the predictors until convergence.

Note that  $\hat\beta$   will typically be sparse, and
so the computation $w=2W_{11} \hat\beta$
will be fast:
if there are $r$ non-zero elements, it takes $rp$ operations.

Finally, suppose our final estimate of $\Sigma$ is $\hat \Sigma=W$,
and store the estimates $\hat\beta$ from the above in the rows
and columns  of a  $p\times p$ matrix $\hat B$ (note that
the diagonal of $\hat B$ is not determined).
Then we can obtain the $p$th row (and column) of $\hat \Theta=\hat \Sigma^{-1}=W^{-1}$ as follows:
\begin{eqnarray}
\hat \Theta_{pp}&=&\frac{1}{W_{pp}-2\sum_{k\neq p} \hat B_{kp}W_{kp} }\cr
\hat \Theta_{kp}&=&-2\hat \Theta_{pp}\hat B_{kp};\; k\neq p
\end{eqnarray}
Interestingly, if $W=S$, these  are just the formulas for obtaining 
the inverse of a partitioned matrix.
That is, if we set $W=S$ and  $\rho=0$ in the above algorithm, then one sweep
through the predictors computes $S^{-1}$, using a linear regression
at each stage.

\section{Timing comparisons}
We simulated Gaussian data from  both {\em sparse} and {\em dense} scenarios,
for a range of problem sizes $p$. The sparse 
scenario is the AR(1) model taken from \citeasnoun{YL2007a}: $\beta_{ii}=1$,
$\beta_{i,i-1}=\beta_{i-1,i}=0.5$, and zero otherwise.
In the dense scenario, $\beta_{ii}=2$,$\beta_{ii'}=1$ otherwise.
We chose the the penalty parameter so that the solution had about the actual 
number of non-zero elements in the sparse setting, and about half of
total number of elements in the dense setting.
The convergence threshold was $0.0001$.
The  covariance lasso procedure was coded in Fortran, linked to an R language function.
All timings  were carried out on a
 Intel Xeon  2.80GH processor.

We compared the covariance lasso 
to the COVSEL program provided by \citeasnoun{BGA2007}. This 
is a Matlab program, with a loop that calls a  C language code to do the box-constrained
QP for each column of the solution matrix. To be as fair as possible
to COVSEL, we only counted the CPU time spent in the C program.
We set the  maximum number of outer iterations to 30, and following
the authors code, set the 
the duality gap for convergence to 0.1.

The number of CPU seconds for each trial is shown in Table \ref{tab:speed}.
\begin{table}
\begin{center}
\begin{tabular}{lcrrrr}
$p$     & Problem&  (1) Covariance& (2) Approx& (3) COVSEL &Ratio of\\
     & Type& Lasso & & &(3) to (1) \\
\hline
 100    & sparse  &       .018 &   .007 & 34.67  & 1926.1\\
 100    &   dense  &     .038  & .018 &  2.17   &  57.1\\
\hline
 200    &   sparse  &    .070  & .027 &  $>205.35$  &$>2933.6$  \\
 200    &   dense    &   .324  &  .146 &16.87 &52.1   \\
\hline
 400   &     sparse  &   .601 & .193&  $>1616.66$ & $>2690.0$\\
 400  &     dense      &  2.47  & .752  & 313.04 & 126.5\\
\end{tabular}
\end{center}
\caption[tab:speed]{\em Timings (seconds) for covariance lasso, Meinhausen-Buhlmann
approximation, and COVSEL procedures.}
\label{tab:speed}
\end{table}
In the dense scenarios for $p=200$ and 400, COVSEL had not converged by
30 iterations.
We see that the covariance Lasso is 50 to 2000 times faster than COVSEL,
and only about 3 times slower than the approximate 
method. Thus the covariance lasso is taking only about 3 passes
through the the columns of $W$ on average.

\begin{figure}
\begin{center}
\epsfig{file=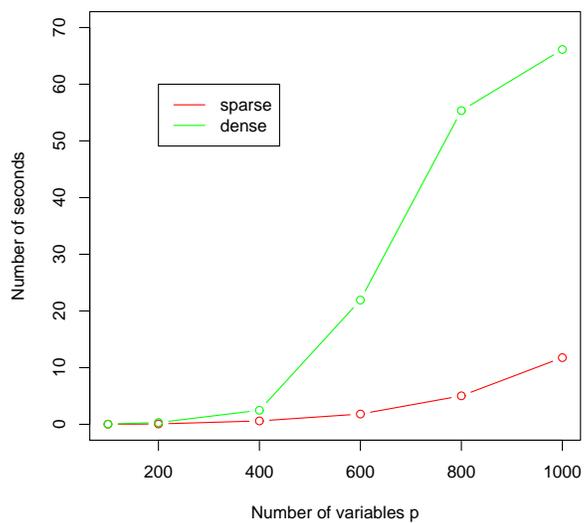,width=.6\textwidth}
\end{center}
  \caption{\small\em Number of CPU seconds required for the
covariance lasso procedure. }
  \label{figspeed}
\end{figure}

Figure \ref{figspeed} shows the number of CPU seconds required for the
covariance lasso procedure, for problem sizes up to 1000.
Even in the dense scenario, it solves a 1000 node problem ($\sim 500,000$ parameters)
 is about a minute.

\section{Analysis of cell signalling data}
\begin{figure}
\begin{center}
\epsfig{file=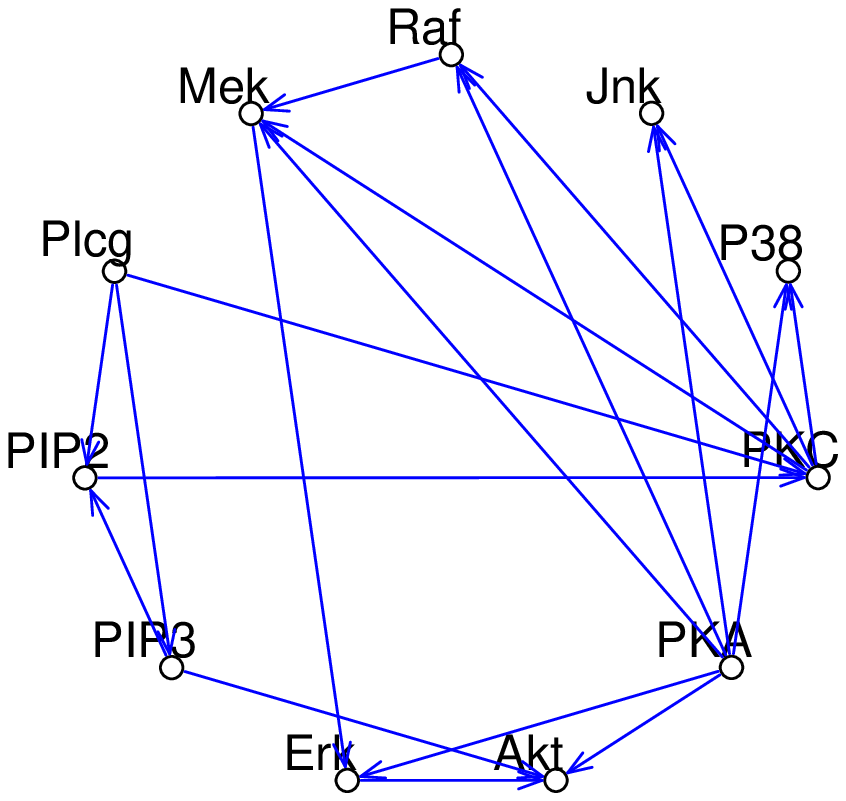,width=.3\textwidth}
\end{center}
  \caption[figsachs0]{\small\em Directed acylic graph from cell-signaling data,
from \citeasnoun{sachs2003}.}
  \label{figsachs0}
\end{figure}

For illustration we 
  analyze a flow cytometry dataset on $p=11$ proteins and  $n=7466$ cells, from
\citeasnoun{sachs2003}. These authors fit a directed acyclic graph (DAG)
to the data, producing the network in Figure \ref{figsachs0}.

\begin{figure}
\begin{center}
\epsfig{file=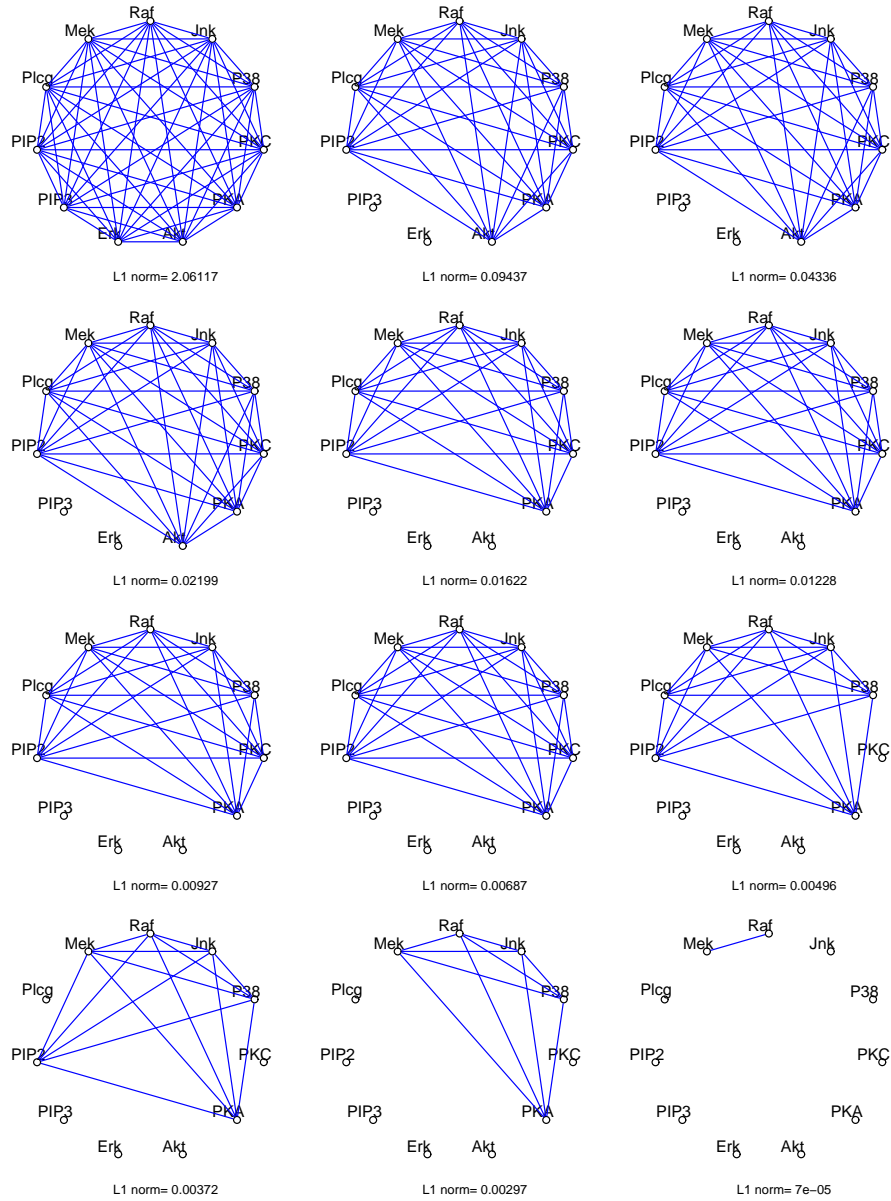,width=.9\textwidth}
\end{center}
  \caption{\small\em Cell-signaling data: undirected graphs from covariance lasso with different values of the penalty parameter $\rho$.} 
  \label{figsachs}
\end{figure}

\begin{figure}
\begin{center}
\epsfig{file=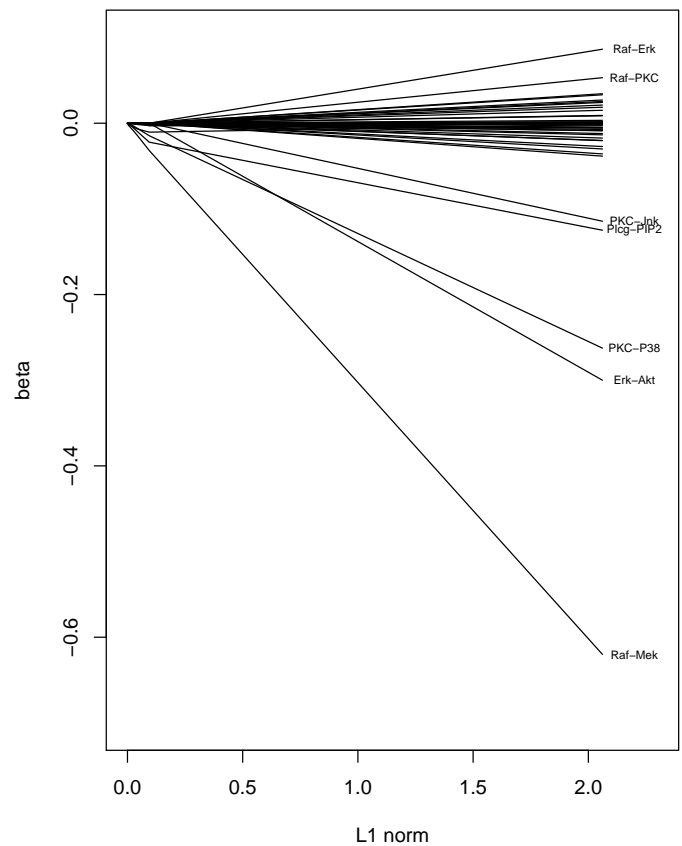,width=.7\textwidth}
\end{center}
  \caption{\small\em Cell-signaling data: profile of coefficients as the
total $L_1$ norm of the coefficient vector increases, that is, as $\rho$ decreases.
Profiles for the largest  coefficients are labeled with the corresponding
pair of proteins.}
  \label{figsachs2}
\end{figure}

The result of applying the covariance Lasso to these data is shown in Figure \ref{figsachs},
for 12 different values of the penalty parameter $\rho$.
There is moderate agreement between, for example, the graph for L1 norm 
 $=0.00496$ and the DAG:
the former has about half of the edges and non-edges that appear in the DAG.
Figure \ref{figsachs2} shows the lasso coefficients as a function of total $L_1$ norm
of the coefficient vector.

\begin{figure}
\begin{center}
\epsfig{file=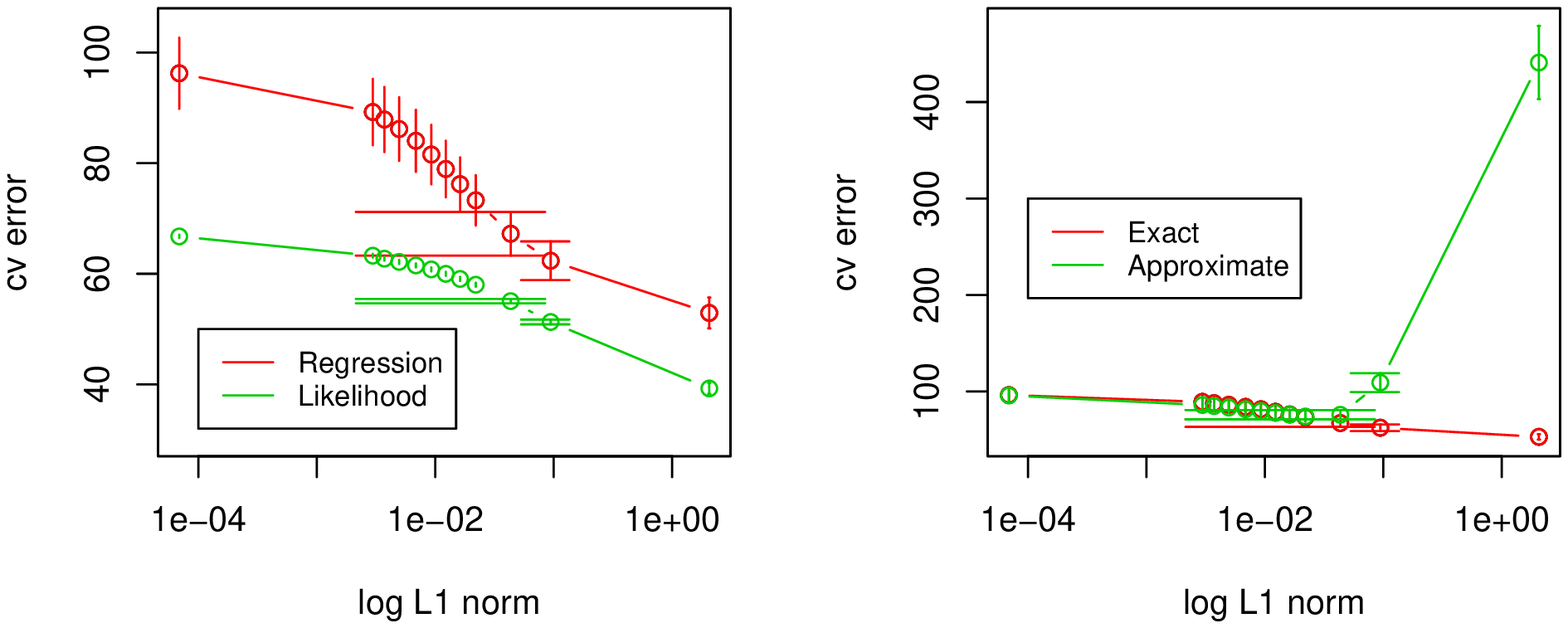,width=\textwidth}
\end{center}
  \caption{\small\em Cell-signaling data. Left panel shows tenfold cross-validation
using both Regression and Likelihood approaches (details in text).
Right panel compares the  regression sum of squares of the
exact covariance lasso approach to the Meinhausen-Buhlmann approximation. }
  \label{figsachs-cv}
\end{figure}

In the left panel of Figure \ref{figsachs-cv} we  tried two different kinds of 10-fold cross-validation
for estimation of the parameter $\rho$.
In the ``Regression'' approach, we fit the covariance-lasso to nine-tenths of the
data, and  used the penalized regression model 
for each protein to predict the value of that protein in the validation set.
We then averaged the squared prediction errors over all 11 proteins.
In the ``Likelihood'' approach, we again applied the covariance-lasso to nine-tenths of the 
data, and then evaluated the log-likelihood (\ref{one}) over the validation set.
The two cross-validation curves indicate that the unregularized
model  is the best, not surprising give the large number
of observations and relatively small number of parameters.
However we also see that the likelihood approach is far less variable than
the regression method.

The right panel compares the  cross-validated sum of squares  of the 
exact covariance lasso approach to the Meinhausen-Buhlmann approximation.
For lightly regularized models, the exact approach has a clear advantage.

\section{Discussion}
We have presented a simple and fast algorithm
for estimation of  a sparse inverse covariance matrix using 
an $L_1$ penalty. It cycles through the variables, fitting a modified  lasso
regression to each variable in turn.
The individual lasso problems are solved by coordinate descent.

The speed of this new procedure
should facilitate  the application of sparse inverse covariance
procedures to large datasets involving thousands of parameters.

 Fortran and R language routines for the proposed methods will be 
made freely available.

\subsubsection*{Acknowledgments}
We thank the authors of \citeasnoun{BGA2007} for making  their
COVSEL program publicly available,
and Larry Wasserman for helpful discussions.
Friedman  was partially supported by grant DMS-97-64431 from  the National
Science Foundation.
Hastie was partially supported by grant DMS-0505676 from the National
Science Foundation, and grant 2R01 CA 72028-07 from the National Institutes of
Health.
Tibshirani was partially supported by National Science Foundation
Grant DMS-9971405  and National Institutes of Health Contract  N01-HV-28183.

\bibliographystyle{agsm}
\bibliography{graph}

\end{document}